**Fundamental problems of modeling the dynamics of internal gravity waves**

**with applications to the Arctic Basin**

**Vitaly V. Bulatov, Yury V. Vladimirov**

**Institute for Problems in Mechanics**

**Russian Academy of Sciences**

**Pr. Vernadskogo 101 - 1, Moscow, 119526, Russia**

**internalwave@mail.ru**

**Abstract**

In this paper, we consider fundamental problems of the dynamics of internal gravity waves. We present analytical and numerical algorithms for calculating the wave fields for a set of values of the parameters, as observed in the ocean. We show that our mathematical models can describe the wave dynamics of the Arctic Basin, taking into account the actual physical characteristics of sea water, topography of its floor, etc. The numerical and analytical results show that the internal gravity waves have a significant effect on underwater sea objects in the Arctic Basin.

## INTRODUCTION

The history of studying the internal gravity waves in the ocean, as is known, originated in the Arctic Region after F. Nansen had described a phenomenon called "Dead Water". Nansen was the first man to observe the internal gravity waves in the Arctic Ocean. The notion of internal waves involves different oceanic phenomena such as "Dead Water", internal tidal waves, large scale oceanic circulation, and powerful pulsating internal waves. Such natural phenomena exist in the atmosphere as well; however, the theory of internal waves in the atmosphere was developed at a later time along with progress of the aircraft industry and aviation technology [1,2].

Studying the oceanic currents of the Arctic Ocean became the principal objective of the Fram expedition in 1893-1986 and was continued in the years to follow. At that time such a voyage was an equivalent of a travel to the Moon. In the process of the expedition the scientists made a lot of observations and collected many data sheets and measurements in the Arctic which had been essentially unexplored at the time .

During his arctic journey F. Nansen was the first scientist to classify the manner in which the "Dead Water" phenomenon occurs. This phenomenon comes about from the internal gravity waves generated by a slow moving vessel. The first theoretical work dedicated to internal gravity waves was the thesis work by V.W. Ekman, who provided a detailed definition of dead water and systematized the data obtained by F. Nansen .

The "Dead Water" effect from internal gravity waves has been long known to sailors. Sailing vessels after being caught in the thermocline (a density contrast layer) suddenly brought down to a complete stop. This phenomenon resulted from the internal gravity waves generated by the vessel. But since the sailors saw no waves on the surface behind the ship this enormous water resistance seemed to be inexplicable whatsoever, and they blamed the bewitched drowned for holding the ship in place and not letting her go.

Up to the 1960-s of the 20 century the research was for the most part focused on tidal waves, however, in the middle of the 1950-s some theoretic developmental studies and laboratory investigations were undertaken that involved the internal pulsating waves. As early as in 1950 there appeared the first definition for a superficial wake of internal gravity waves in the ocean. In 1965 the first scientific observations were made concerning the oceanic large amplitude internal waves and solitons.

The interest to investigations involving the internal gravity waves grew up after the WW2 when the US Navy lost a few of its most advanced at the time submarines. After those accidents there were assumptions made that the disaster might have been caused by the internal gravity waves. As is known, the submarines often move along the thermocline (a density contrast layer) to avoid detection since the thermocline surface reflects the acoustical signals of active sonars and sea vessels .

The most notorious incident involved the US Navy Thresher submarine that was lost at sea in 1963 with the crew of 129 on board. The US Thresher submarine was a most advanced boat in the world in the 1960-s and she could descend to depths and move at velocities that were inconceivable just a few years before she was constructed. It might be that the Thresher submarine was going along the thermocline and a large internal wave took her down to a depth pressure that she could not survive. There were no failures reported in operation of the submarine instrumentation, and no severe storms were detected in the area where the submarine was lost. It all might happen very quickly since the crew was not able to prevent the boat from falling down to deep water

The first scientific explanation of what might be happening with the submarines appeared in 1965. It was the year that in the Andaman Sea for the first time ever discovered were large internal waves which happened to be a real sensation. Moving along the thermocline the wave could go 80 meters down. The oceanographers in the world until then believed no such waves existed. However, the Russian and US space programs allowed the scientists to take a look at our planet from the space. The panoramic photographs made from the orbit showed multiple wakes of waves. The point is that internal waves can create rather strong currents on the ocean surface. The flow is changed depending on the wave extension: its velocity is greater at the wave crest and wave trough, and is slower where thermocline oscillations are little. If several wave packets follow each other this pattern on the ocean surface is repeated. These surface flows are getting stronger or weaker when affected by wind waves depending on the set of wind, and can be defined as variations in light reflecting capacity of the ocean surface by remote radar sensing [1-3].

The Apollo-Soyuz Test Project in 1975 was the first joint Russian-US enterprise in the space. The NASA researchers asked the crew to monitor the internal waves and photograph them. John Apel, who was a pioneer in studying the internal waves of the World Ocean, in 1978, wrote in the general scientific report of the expedition the following :

«At least three photographs made by the "Apollo-Soyuz" crewmen have revealed obvious signs of internal gravity waves in the ocean, which is evident from periodically changing optical reflections from the ocean surface positioned above those waves. The wave packet (or the wave group) observed at Cadiz in Spain had the characteristics similar to those of internal waves shown by satellite photos taken close to the East Coast of the USA. In the Andaman Sea near the Malay Peninsula observed were several wave groups with the wave-lengths of 5 to 10 km and separations between the groups of 70 to 115 km. If these are really surface wakes of internal waves, these waves are one of the largest and fastest to this day. The measurements made earlier from aboard a sea vessel indicated presence in that area of large amplitude internal waves».

The destruction of submarines gives evidence of the force of internal gravity waves. The internal waves generally move along the thermocline (a density contrast layer) positioned at a certain depth which separates by rather a weak at the ocean surface from deep waters, and their oscillation vector is directed either downwards or upwards. Once occurred these waves are propagating while maintain their form and force, and are capable of covering long distances. The internal waves also function as a carrier vehicle by transferring biomass and nourishment from one place to another. The underwater waves traveling upwards the shelf take the nourishment from the ocean deep water to the more salty shallow waters with ideal living conditions for larvae and fingerling. The wave motion in this case may be compared to a pumping action .

The amplitude of internal gravity waves is generally comparable to the depth of the near-surface ocean. However, there was reported an occurrence when the wave was five times higher than the thermocline height. Since the sea water always contains layers positioned above each other with different temperature and salinity characteristics the internal gravity waves are generally in existence everywhere within the ocean thickness, but reach their maximum amplitudes typically near the thermocline. In equatorial areas the thermocline is located at the depth of 200 to 300 m, in the region of the Ormen Lange gas-field (Norway, Arctic basin) it is at 550-m depth, and in the Norwegian fiords with flowing in fresh water the thermocline is just 4 to 10 m deep .

The industrial activities on the continental shelf involving crude-oil and gas production and other mining works have become an important factor for beginning the research on internal gravity waves with large amplitudes. The vessels and rigs for drilling and underwater constructions use long tubes connecting them to the sea bottom. The builders of underwater structures in equatorial areas have experienced the effect of large underwater waves and strong surface flows that can be shaped as a steep waterfall. Some time ago, when the phenomenon of internal wave was not known yet there were times when the builders got their equipment lost. Such losses are quite costly and make it clear that to protect and keep safe the fixed structures at sea we have to control the effect of internal gravity waves [1,2].

The construction of sea platforms such as, for example, the Ormen Lange gas-field (Norway, Arctic basin) and other constructions at the sea bottom have stipulated many scientific studies including the fundamental research. Thus, for instance, the thermocline at the Ormen Lange gas-field (Norway, Arctic basin) is located at the depth of 500 m. It separates the Atlantic warm water of some $7^0$C from the polar cold water of about $1^0$C. The additionally accumulated warm current of the Atlantic Ocean can drop the thermocline even lower. The measurements in the region of Ormen Lange have registered once to lower the thermocline down from its regular depth to 550 m where it stayed for three days. It went down to the platform positioned at 850-m depth. After that the water was flowing back and upslope. In the beginning its motion velocity was half a meter per second which was very fast for a near-bottom current. Gradually the velocity dropped down, but the oscillations continued for a surprisingly long time of full 24 hours .

The special interest to the research involving internal gravity waves is attributed also to intensive development of the Arctic and its natural resources. The internal waves are still poorly studied in the Arctic region since they are moving below the ice and practically are not visible from above. However, the available information on the movement of underwater objects indicates their presence. Yet, there may be exceptions when the internal gravity waves reach the ice cover lifting it up or down with certain periodicity, which can be monitored radar sensing equipment. The effect of waves of all types can result in breaking the Arctic ice cover. In addition the waves provoke iceberg displacement and move various pollutants. This is why the research of wave dynamics in the Arctic shelf region appears to be an important scientific and practical task to ensure safety in construction and operation of sea platforms .

To make a detailed description of a wide range of physical phenomena that belong to wave dynamics of stratified, horizontally non-uniform and non-stationary mediums one should proceed from rather advanced mathematical models which usually become quite complex non-linear and multivariate, and can be fully and effectively explored only if using numerical methods. In certain situations, however, an adequate initial representation of the explored phenomena circle can be obtained when using more simple asymptotic models and analytical methods. For that matter there are as quite characteristic the problems of mathematically modeling the dynamics for non-harmonic packets of internal gravity waves, and even within the bounds of linear models they offer rather specific solutions that provide along with nontrivial physical effects for a self-sustained mathematical interest

Now in connection with  the new problems arising in geophysics, oceanology, physics of atmosphere, usage of the cryogenic liquids in the engineering sphere, as well as the problems of protection and study of the medium, operation of the complex hydraulic engineering facilities, including  the marine oil producing complexes, and  a number of other actual problems facing the science and engineering we can observe the growth of interest to the research of the dynamics of the wave movements of  the different inhomogeneous liquids and, in particular, the stratified liquids. This interest is caused not only by the practical needs, but also by the need to have the solid theoretical base to solve the arising problems [1-3].

It is necessary to note, that solution of the problems of the mechanics of continua and hydrodynamics always served as the stimulus of new directions in mathematics and mathematical physics. As the illustration to the above may serve the stream of the new ideas in the theory of the nonlinear differential equations, and also the discovery of the startling dependencies between the can be appearing  the different  branches of mathematics, that has followed after exploration of Cartevega de Vriza  equation for the waves on the shallow water. Certainly, for the detailed description of the big amount of the natural phenomena connected with the dynamics of the stratified non-uniform in the horizontal direction and the non-stationary mediums, it is necessary to use the sufficiently developed mathematical models, which as a rule are the rather complex nonlinear multiparametric mathematical models and for their full-size research only the numerical methods are effective.

The interest to the internal gravity waves is caused by their wide presence in the nature. Both the air atmosphere, and the oceans (Arctic basin) are stratified. Reduction of the air pressure and its density at the increase of the elevation are well known. But the sea water is also stratified. Here the raise of the water density with the increase of its depth is determined, mainly, not by the rather small compressibility of the water, but by the fact, that with the increase of the depth, as a rule, the temperature of the water is decreasing, and its saltiness grows. In the capacity of the stratified medium, as a rule, one considers the medium, the physical characteristics (density, dynamic viscosity and others) of which in the medium stationary status are changing only along some concrete direction. Stratification of the natural mediums (the ocean, the atmosphere) can be caused by the different physical reasons, but the most often by the gravity. This force creates in the stratified medium such a distribution of the particles of the dissolved in it salts and suspensions, at which it forms the heterogeneity of the medium along the direction of the gravity field in the stratified medium.

This heterogeneity is called the density stratification. The stratification of density, as the experimental and natural observation show, renders the most essential influence, as compared with other kinds of stratification, on the dynamic properties of the medium and on the processes of distribution in the medium of the wave movements. Consequently at consideration of the wave generations in the stratified mediums usually neglect all other kinds of wave stratification, except for the density stratification, and in the capacity of the stratified medium they consider the medium with density stratification caused by the gravity.

In the real oceanic conditions (Arctic basin) the density changes are small, the periods of oscillations of the internal waves are changing from several minutes (in the layers with rather fast change of the temperatures and the depth) up to the several hours. Such great periods of the fluctuations means, that even at the big amplitude of the internal waves, but they can achieve dozens of meters along the vertical direction , the speeds of the particles in the internal wave are low – for the vertical components the speeds of the particles have the order of mm/s, and for the horizontal - cm/s. Therefore the dissipative losses – the losses caused by of the liquid viscosity in the internal waves are very small, and the waves propagation can propagate practically without fading within the big distances,. At that the speed of propagation of the internal waves in the ocean is low - the order of dozens of cm/s .

These properties of the internal gravity waves mean, that they can keep the information about the sources of their generations for the long time. Unfortunately, it is very difficult to orientate in this information because the internal waves pass the dozens and hundreds of kilometers from the source the generations up to the place of supervision; and practically everywhere, where there is the stratification of the ocean takes place, we can observe the internal waves, but simultaneously we can "hear" the "voices" of the most different sources. At that the qualitative (and the quantitative) properties of the internal waves, caused by that or other concrete source depend not only on its physical nature, and also on its spatial and time distribution, but also depends on the properties of the medium located between the source of the waves and the place of the observation .

The internal waves represent the big interest not only from the point of view of their applications. They are of the interest to the theorists occupied with the problem of propagation ша the waves, as the internal waves properties in many respects differ from the properties of the accustomed to us the acoustic or electromagnetic waves. For example, for short harmonic internal waves of the following kind $A \exp(ikS(x, y, z) - i\omega t)$ , where $k \gg 1$ ) – the rays are directed not perpendicularly to the wave fronts – to the surfaces of the equal phase $S = const$ , but along these surfaces .

The stratification, or the layered structure of the natural mediums (oceans and the air atmosphere) causing formation of the internal gravity waves plays then appreciable role in different oceanic and atmospheric processes and influences on the horizontal and vertical dynamic exchanges. The periods of the internal waves can make from several minutes up to several hours, the lengths of the waves can to achieve up to dozens of kilometers, and their amplitudes can exceed dozens of meters. The physical mechanism of formation of the internal waves is simple enough: if in the steadily stable stratified medium has appeared a generation, which has caused the particle out its balance state, then under action of gravity and the buoyancy the particle will make fluctuations about its balance position .

The theory of the wave movements of the stratified mediums being the section of the modern hydrodynamics is quickly developing recently and rather interesting in the theoretical aspect as well as it is connected with the major applications in the engineering field (hydraulic engineering, shipbuilding, navigation, energy) and in geophysics (oceanology, meteorology, hydrology, preservation of the environment). Now the majority of the applied problems, concerning the waves generation caused by various generations are solved just in the linear aspect, that is considering the assumption, that the amplitude of the wave movements is small in comparison with length of the wave. The relative simplicity of the solution of the linear equations as compared with the solution of the complete nonlinear problem, the modern development of the corresponding mathematical tools and the computer engineering allows to meet many challenges of practice .

Initially the theory of wave movements of the stratified medium was developing as the theory of superficial waves describing the behavior of the free surface of the liquid being in the gravity field. Later it has been understood, that the superficial waves represent the special type of the waves existing on the border of the separation of the various mediums densities, which in turn represent the special case of the internal waves in the medium non-uniform (stratified) in density. In the real ocean (Arctic basin) the non-uniform distribution of density may take place both in the vertical, and in the horizontal directions. At that considering the existing heterogeneity of the medium both in the vertical the horizontal directions, and also its nonstationarity at research of the distribution of the internal gravity waves require to use the special mathematical tools. As a rule it is supposed, that the density distribution is steady, that is the density does not decrease with the change of the depth .

The reasons of initiation of the superficial and internal waves in the real ocean are very different: the fluctuations of the atmospheric pressure, the flow past of the bottom asperities, movement of the surface or the underwater ship, deformation in the density field, the turbulent spots formed by any reasons, the bottom shift or the underwater earthquake, the surface or underwater explosions, etc. One of the mechanisms of generation of the internal gravity waves may be excitation of the wave fields caused by, for example, at movement (flow past) of the non-local sources (underwater vessels, sea platforms), the turbulent spots, the water lenses and the other non-wave formations with the abnormal characteristics.

## 1. PROBLEM FORMULATION

*1.1. Wave dynamics in vertically stratified mediums*. Generally the system of the linear equations describing the small movements of the originally quiescent incompressible non-viscous stratified medium in the system of the Cartesian coordinates x = (x, y, z) with the axis z directed vertically upwards, looks like [1,2]

div U = Q (x, t)

$$\rho_0 \frac{\partial U}{\partial t} + grad\,p + F = S(x,t) \qquad\qquad (1)$$

$$\frac{\partial \rho}{\partial t} + \frac{d\rho_0}{dz} W = K(x,t)$$

where U = $(U_1, U_2, W)$, p, $\rho$ - perturbation of the velocity vector, pressure and density; $\rho_0$(z) – stratified medium density in the quiescent state; F=(0,0,g$\rho$), g - acceleration of the gravity. Functions Q, S, K represent intensities of distributions of the sources of weight, pulses and density accordingly. Boundary conditions on the free surface $z = 0$ and on the flat bottom $z = -H$ look like

$W = \partial\eta / \partial t$   p - g$\rho_0\eta$ = P(x, y, t)      $z = 0$

$\qquad\qquad (2)$

$W = Z(x,y,t)\ \ z = -H$

Here function η (x, y, t) describes the vertical displacement of the free surface; P - external pressure, acting on the free surface; and Z – the vertical speed of the bottom. The initial conditions at t=0 are as follows:

$U = U^*(x)$ ,ρ=ρ*(x),  η=η*(x, y)      $\qquad\qquad (3)$

where functions $U^*(x)$, ρ*, η* - initial values of generations of the vector of speed, density and elevation of the free surface. To ensure the correct performance of the condition it is required to meet the following condition: div $U^*(x)$ = Q ($t = 0$).(t=0).

By virtue of the linearity of the problem the forced waves are represent  by the superposition of the free harmonious waves described by the homogeneous system (1) and the homogeneous boundary and initial conditions of (2), (3). The system (1) can be reduced to one equation for any of required functions, usually it is done for the vertical component of speed. At that the homogeneous system (1) and the homogeneous boundary conditions (2) may be presented in the form

$$\frac{\partial^2}{\partial t^2}\left((\frac{\partial^2}{\partial z^2}+\Delta)W - \frac{N^2(z)}{g}\frac{\partial W}{\partial z}\right) + N^2(z)\Delta W = 0$$

$$\frac{\partial^3 W}{\partial z \partial t^2} - g\Delta W = 0, z = 0 \qquad\qquad (4)$$

$$W = 0, -H$$

$$\Delta = \frac{\partial^2}{\partial x^2} + \frac{\partial^2}{\partial y^2}, \ N^2(z) = -g\frac{d}{\rho_0}\frac{\rho_0}{dz}$$

The first equation from (4) is to some extend simplified after introduction of Boissinesq approximation. At usage of this approximation in the equations of the pulse preservation (1) the difference of the density from some constant value $\rho_s$, is considered only in the member describing floatability, in the inertial members the real density is replaced with the value $\rho_s$, and the equation (4) is reduced to the kind:

$$\frac{\partial^2}{\partial t^2}(\frac{\partial^2}{\partial z^2}+\Delta)W + N^2(z)\Delta W = 0 \qquad\qquad (5)$$

The function $N(z)$ is one of the basic characteristics of the stratified medium, and has the fundamental value in the theory of the internal gravity waves and is called the buoyancy frequency or Vaisala-Brunt frequency. The value T=2π/N defines Vaisala-Brunt period. For the real ocean and the atmosphere the value T varies from minutes up to several hours, and for the stratified liquid produced in the laboratory, it can make some seconds .

Homogeneity of the equations (4), (5) and their boundary conditions at the variables x, y, t allow to look for the elementary wave solutions in the field of the plane waves: $W(x,t) = \varphi(z)\exp(ikr - i\omega t)$, where k is the wave vector in the plane x, y; $\omega$ - oscillations frequency; r = (x, y).

For function $\varphi(z)$ from (4) the boundary problem results in the following Sturm-Liouville equation:

$$\frac{\partial^2 \varphi}{\partial z^2} - \frac{N^2(z)}{g}\frac{\partial \varphi}{\partial z} + \left(\frac{N^2(z)}{\omega^2} - 1\right)k^2\varphi = 0 \qquad\qquad (6)$$

$$\frac{\partial\varphi(0)}{\partial z} = g\ \kappa^2\varphi(0)/\omega^2, \quad \varphi(-H) = 0,$$

in Boissinesq approximation

$$\frac{\partial^2 \varphi}{\partial z^2} + \left( \frac{N^2(z)}{\omega^2} - 1 \right) \kappa^2 \varphi = 0 \qquad (7)$$

where k = | k |. Problems (6), (7) are the problems of the own values, after solution of which , there may be defined the system of the own values $\omega$ (dispersive dependences) and own functions $\varphi(z)$ for each fixed value of the wave number k. The spectrum of such problems is always discrete, that is the system possesses the countable number of the modes $\varphi_n(z)$ (n = 1,2,3...), to each of which there corresponds the law of the dispersion $\omega_n = \omega_n(\kappa)$. In the case when the depth of the liquid is endless and the difference of the function N (z) from zero takes place also within the unlimited interval, then alongside with the discrete spectrum there is also a continuous spectrum.

The knowledge of the dispersive dependences and their properties has the paramount value at research of the linear gravity waves. The basic properties of the own values and the own functions of the problems (6), (7) are well studied. The own functions of the considered problems may be divided into to two classes. The first class is presented by one own function $\varphi_0(z)$, which is monotonically and quickly enough decreasing with the increasing depth. This own function poorly depends on the conditions of stratification and describes the superficial wave. All other own functions correspond to the normal modes of the internal waves. For the internal waves own function $\varphi_n(z)$ ( n=1,2,3...) has n-1 zero inside of the interval [-H, 0]. For the continuously stratified liquid of the final depth both for its superficial wave and for its internal waves is typical the monotonous increase in frequency ω of a single mode at the growth of the wave number k, the monotonous reduction of the phase speeds $c_f = \omega / \kappa$ with the growth of k and at the increase of the mode number, and also excess of the phase speeds over the group speeds $c_g = d\omega / d\kappa$. The maximal values of the phase and the group speeds coincide and take place at k = 0. The significant difference of the superficial wave from the internal waves consists that in the short-wave region (κ→∞) the frequency of the superficial wave is unrestrictedly increasing ($\sim \kappa^{1/2}$), whereas the internal waves frequency tends to the value $\max_z N(z)$.

Rather small change of the liquid density at changing the depth in comparison with the drop of the density on the water – air border allows to research the internal waves in the approximation of the "solid cover" ($\varphi(0) = 0$), which filters the superficial waves out without essential distortion of the internal waves. Approximation of "the solid cover" allows to neglect the first sum component in the dynamic condition of (2).

The analytical decision of the problems (6), (7) is possible only for some special cases of changing of N(z) function. At the smooth changing of the function N(z) the WKBJ approximation method is frequently applied for the approximate calculation of the own values and the own functions. However this approach is limited by the case, when the function N(z) has no more than one maximum [1-4].

More accurate results may be received by direct use of the numerical methods, and at the present tome there are several methods of the numerical solution of the problems (6), (7): the finite-difference approximation method, at which the differential equations (6), (7) and the boundary conditions are replaced with the system of the difference equations, approximation of the initial continuous distribution of density of the piecewise-constant function. In this case there is a possibility of existence of only the final number of the wave modes. The analysis of the asymptotic behavior of the phases velocities $c_f$ in the shortwave field has demonstrated, that in the stratified medium with the step-by-step stratification $c_f \approx k^{-1/2}$, while for the medium with the continuous profile of density $c_f \approx k^{-1}$. Piecewise constant approximation of the Vaisala-Brunt frequency. The numerical solution of the differential equations, derived from (6), (7) after introduction of the Prewfer modified transformations

$$\varphi(z) = \exp(az)\sin b(z) \qquad (8)$$

$$\frac{d\varphi(z)}{dz} = \exp(az)\cos b(z)$$

As a result of the transformations (8) for definition of the dispersive dependences it is enough to solve the nonlinear boundary value problem of the first order for the function $b(z)$, behavior of which unlike $\varphi(z)$ is monotonous [2, 5,6].

The up to now cumulative experience of calculation of dispersive dependences demonstrates, that their most complex behavior arises at the presence in the stratified medium of several wave guides and on the charts of the dispersive curves there may arise the nodes and crowdings, which testify, that the behavior of the group speeds of the internal waves becomes non-monotonic and on some (abnormal) frequencies the different modes extend practically with the identical phase speeds, having the different group speeds. Such areas are called the resonant zones and in them conditions for an overflow of the energy from the lowest energy-carrying modes into the highest energy-carrying modes are created. This phenomenon looks like as insignificant in application to the linearized problem, but may be important at considering the nonlinear members. The abnormal frequencies represent the rather important feature of the internal waves, on them there is a qualitative change of the vertical structure of the wave field .

The thin structure of distribution of the Vaisala-Brunt frequency also may bring to the similar effects of the crowding of the dispersive characteristics on depth. The dispersive curves under action of the thin hydrological structure can be stratified into the separate groups (clusters) inside which occurs the rapprochement of the dispersive parameters of the different mode, whereas the groups themselves are moving away from each other. Such stratification, apparently, may affect on the spectra of the internal waves in the field of the frequencies close to the maximum Vaisala-Brunt frequency.

For the solution of the equations (1) with conditions of (2), (3) rather convenient method of solution is application of Green functions describing development of generations caused by an instant dot source, being on the depth of $z_1$. In case of the system homogeneous in the horizontal direction it is useful to use Fourier expansion

$$G(\mathbf{r}, z, z_1, t) = \int \frac{d\mathbf{k}}{(2\pi)^2} \frac{d\omega}{2\pi} e^{i(\mathbf{k} \cdot \mathbf{r} - \omega t)} G(\mathbf{k}, z, z_1, \omega)$$

(9)

Then $G(\mathbf{k}, z, z_1, \omega)$ should satisfy the equation of the following kind:

$$LG(\mathbf{k}, z, z_1, \omega) = \delta(z - z_1)$$ 

(10)

$$L = \frac{\partial}{\partial z} \rho_0(z) \frac{\partial}{\partial z} + \rho_0 k^2 \left( \frac{N^2(z)}{\omega^2} - 1 \right)$$

The solution of this equation one should look for in the form of the eigenfunctions expansion of the problem (6)

$$G(\mathbf{k}, z, z_1, \omega) = G_0(\mathbf{k}, z, z_1) + \sum_n \frac{\omega_n^2(k) \varphi_n(k, z) \varphi_n(k, z_1)}{\omega^2 - \omega_n^2(k)}$$

where $G_0(\mathbf{k}, z, z_1)$ is the solution of the equation (10) at $\omega \to \infty$ and describes an instant part of the medium response to the external excitation, and the sum of the eigenfunctions describes the contribution of the wave part. Usually the value $G_0$ is rejected without any discussions. However in some cases, for example, at calculation of the amplitudes of the waves from the periodic sources, this component may be essential, because for the internal waves the law of decrease of the amplitudes of the wave and the non- wave parts of the excitation as the distance from the source of excitation increases is identical [1-3].

At fulfillment of the inverse Fourier transformation there is an ambiguity connected with the necessity to set the rule for the flow past the singularities on the real axis $\omega$. The choice of the unambiguous solution is achieved at imposing the causality requirement being reduced to the condition $G(t)\big|_{t<0} \equiv 0$ (Green's retarded function). Green's retarded function corresponds to the solution satisfying the principle of Mandelshtam radiation, when the energy expands from the source. By virtue of the specific law of dispersion of the internal waves the Mandelshtam radiation condition sometimes does not coincide with the Zommerfeld radiation condition (the waves leaving the source), but the use of the Zommerfeld radiation principle at the choice of the unambiguous solution can lead to the incorrect results [1,2,4].

One more method of the choice of the unambiguous solution is the method attributed to Relay providing for introduction of the infinitesimal dissipation equivalent to the Mandelshtam condition. Often the additional condition is set in the form of the requirement of absence of the wave excitations in the distant area upwards the stream (Long condition) however the universality of this condition is not obvious at considering the effect of blocking observable in the stratified liquids. It is also possible to use the approach, at which the stationary solution is considered as a limit at $t \to \infty$ of the non-stationary solution for the acting in the stream source of excitation with the constant characteristics, and which is put into operation $t = 0$.

Let us underline, that the causality condition for Green's function is equivalent to the requirement of analyticity of its transient Fourier transformation in the upper half-plane of the complex frequencies $\omega$. It means, that the features on the real axis should be flow past from above, or in accordance with Feynman rule, to exercise the substitution $\omega \to \omega + i\varepsilon$ $(\varepsilon \to +0)$, having shifted the features from the real axis downwards. The analyticity of the transient Fourier transformation in the upper half-plane $\omega$ enables to write the Cramers-Cronig ratios expressing relationship between the real and the imaginary parts of Green function, and also in the case of $N = const$ by simple way to construct Green function by means of the analytical continuation from the "non-wave" field of $\omega^2 > N^2$ into the "wave" field of $\omega^2 < N^2$ ( the fields, where the equation of the internal waves belongs accordingly to the elliptic or the hyperbolic type).

*1.2 Wave dynamics in horizontally inhomogeneous mediums.* As is well known, an essential influence of the propaganda of internal gravity waves in stratified natural mediums (Arctic basin) is caused by the horizontal inhomogeneity and non-stationarity of these media. To the most typical horizontal inhomogeneities of a real ocean one can refer the modification of the relief of the bottom, and inhomogeneity of the density field, and the variability of the mean flows. One can obtain an exact analytic solution of this problem (for instance, by using the method of separation of variables) only id the distribution of density and the shape of the bottom are described by rather simple model functions. If the shape of the bottom and the stratification are arbitrary, then one can construct only asymptotic representation of the solution in the near and far zones; however, to describe the field of internal waves between these zones, one needs an accurate numerical solution of the problem [1-3].

Using asymptotic methods, one can consider a wide class of interesting physical problems, including problems concerning the propagation of non-harmonic wave packets of internal gravity waves in diverse non-homogeneous stratified media under the assumption that the modification of the parameters of a vertically stratified medium are slow in the horizontal direction. From the general point of view, problems of this kind can be studied in the framework of a combination of the adiabatic and semi-classical approximations or by using close approach, for example, ray expansions. In particular, the asymptotic solutions of diverse dynamical problems can be described by using the Maslov canonical operator, which determines the asymptotic behavior of the solution, including the case of neighborhoods of singular sets composed of focal points, caustics, etc. The specific form of the wave packet can be finally expressed by using some special functions, slay, in terms of oscillating exponentials, Airy function, Fresnel integral, Pearcey-type integral, etc. The above approaches are quite general and, in principle, enable one to solve a broad spectrum of problems from the mathematical point of view; however, the problem of their practical applications and, in particular, of the visualization of the corresponding asymptotic formulas based on the Maslov canonical operator is still far from completion, and in some specific problems to find the asymptotic behavior whose computer realization using software of Mathematica type is rather simple. In this paper, using the approaches developed in [1,2,6,7], we construct and numerically realize asymptotic solutions of the problem, which is formulated as follows.

If we examine the internal gravity waves dynamics for the case when the undisturbed density field $\rho_0(z, x, y)$ depends not only on the depth $z$, but on the horizontal coordinates $x$ and $y$, then, in general terms, if the undisturbed density is a function of horizontal coordinates, such a distribution of density induces a field of horizontal flows. These flows, however, are extremely slow and in the first approximation can be neglected. So it is commonly supposed that the field $\rho_0(z, x, y)$ is defined a priori, thus, it is assumed that there exist certain external sources or the examined system is nonconservative. It is also evident that if the internal gravity waves are propagating above an irregular bottom there is no such a problem, because the "internal wave–irregular bottom" system is conservative and there is no external energy flush .

Then we investigated the following liberalized system of equations of hydrodynamics [1,2]

$$\rho_0 \frac{\partial \tilde{U}_1}{\partial t} = -\frac{\partial p}{\partial x}$$

$$\rho_0 \frac{\partial \tilde{U}_2}{\partial t} = -\frac{\partial p}{\partial y}$$

$$\rho_0 \frac{\partial \tilde{W}}{\partial t} = -\frac{\partial p}{\partial z} + g\rho \qquad\qquad (11)$$

$$\frac{\partial \tilde{U}_1}{\partial x} + \frac{\partial \tilde{U}_2}{\partial y} + \frac{\partial \tilde{W}}{\partial z} = 0$$

$$\frac{\partial \rho}{\partial t} + \tilde{U}_1 \frac{\partial \rho_0}{\partial x} + \tilde{U}_2 \frac{\partial \rho_0}{\partial y} + \tilde{W} \frac{\partial \rho_0}{\partial z} = 0$$

Here $(\tilde{U}_1, \tilde{U}_2, \tilde{W})$ is the velocity vector of internal gravity waves, $p$ and $\rho$ are the pressure and density perturbations, is the acceleration of gravity ( $z$-axis is directed downwards).

Using the Boussinesq approximation which means the density $\rho_0(z, x, y)$ in the first three equations of the system (11) is assumed a constant value, the system (11) by applying the cross-differentiating will be given as

$$\frac{\partial^4 \tilde{W}}{\partial z^2 \partial t^2} + \Delta \frac{\partial^2 \tilde{W}}{\partial t^2} + \frac{g}{\rho_0} \Delta (\tilde{U}_1 \frac{\partial \rho_0}{\partial x} + \tilde{U}_2 \frac{\partial \rho_0}{\partial y} + \tilde{W} \frac{\partial \rho_0}{\partial z}) = 0$$

$$(12)$$

$$\frac{\partial}{\partial t}(\Delta \tilde{U}_1 + \frac{\partial^2 \tilde{W}}{\partial z \partial x}) = 0 \quad , \quad \frac{\partial}{\partial t}(\Delta \tilde{U}_2 + \frac{\partial^2 \tilde{W}}{\partial z \partial y}) = 0$$

As the boundary conditions we take the "rigid-lid" condition: $W = 0$ at $z = 0, H$. Consider the harmonic waves $(\tilde{U}_1, \tilde{U}_2, \tilde{W}) = \exp(i\omega t)(U_1, U_2, W)$. Introduce the non-dimensional variable according to the formulas: $x^* = \frac{x}{L}$, $y^* = \frac{y}{L}$, $z^* = \frac{z}{h}$, where $L$ is the typical scale of the horizontal variations $\rho_0$; $h$ is the typical scale of the vertical variations $\rho_0$ (for example, the thermocline width) .

In non-dimension coordinates the equation system (12) will be written as (index $*$ is omitted hereafter)

$$-\omega^2(\frac{\partial^2 W}{\partial z^2} + \varepsilon^2 \Delta W) + \varepsilon^2 \frac{g_1}{\rho_0}(\varepsilon U_1 \frac{\partial \rho_0}{\partial x} + \varepsilon U_2 \frac{\partial \rho_0}{\partial y} + W \frac{\partial \rho_0}{\partial z}) = 0$$

$$\tag{13}$$

$$-\omega^2(\frac{\partial^2 W}{\partial z^2} + \varepsilon^2 \Delta W) + \varepsilon^2 \frac{g_1}{\rho_0}(\varepsilon U_1 \frac{\partial \rho_0}{\partial x} + \varepsilon U_2 \frac{\partial \rho_0}{\partial y} + W \frac{\partial \rho_0}{\partial z}) = 0$$

$$\varepsilon \Delta U_1 + \frac{\partial^2 W}{\partial z \partial x} = 0 \;,\; \varepsilon \Delta U_2 + \frac{\partial^2 W}{\partial z \partial y} = 0$$

$$\varepsilon = \frac{h}{L} << 1 \;,\; g_1 = \frac{g}{h}.$$

The asymptotic solution (14) shall be found in the form usual for the geometric optics method

$$\mathbf{V}(z,x,y) = \sum_{m=0}^{\infty} (i\varepsilon)^m \mathbf{V}_m(z,x,y) \exp(\frac{S(x,y,t)}{i\varepsilon})$$

$$\tag{14}$$

$$\mathbf{V}(z,x,y) = (U_1(z,x,y), U_2(z,x,y), W(z,x,y))$$

Functions $S(x,y,t)$ and $\mathbf{V}_m$, $m = 0,1,...$ are subject to definition. From here on we shall restrict ourselves to finding only the dominant member of the expansion (15) for the vertical velocity component $W_0(z,x,y)$, at that from the last two equations (13) we have

$$U_{10} = -\frac{i\partial S / \partial x}{|\nabla S|^2} \frac{\partial W_0}{\partial z}, U_{20} = -\frac{i\partial S / \partial y}{|\nabla S|^2} \frac{\partial W_0}{\partial z}$$

$$\tag{15}$$

$$|\nabla S| = \left(\frac{\partial S}{\partial x}\right)^2 + \left(\frac{\partial S}{\partial y}\right)^2$$

Substitute (14) into the first equation of the system (13) and set equal the members of the order O(1)

$$\frac{\partial^2 W_0}{\partial z^2} + |\nabla S|^2 \left(\frac{N^2(z,x,y)}{\omega^2} - 1\right) W_0 = 0$$

$$\tag{16}$$

$$W_0(0,x,y) = W_0(H,x,y) = 0$$

where $N^2(z, x, y) = \dfrac{g_1}{\rho_0} \dfrac{\partial \rho_0}{\partial z}$ is the Vaisala-Brunt frequency depending of the horizontal coordinates.

The boundary problem (16) has a calculation setup of eigenfunctions $W_{0n}$ and eigenvalues $K_n(x, y) \equiv |\nabla S_n|$, which are assumed to be known. From here on the index $n$ will be omitted while assuming that further calculations belong to an individually taken mode.

For the function $S(x, y)$ we have the eikonal equation

$$\left(\frac{\partial S}{\partial x}\right)^2 + \left(\frac{\partial S}{\partial y}\right)^2 = K^2(x, y) \qquad (17)$$

Initial conditions for the eikonal $S$ for the horizontal case are defined on the line $L : x_0(\alpha), y_0(\alpha) :$ $S(x, y)\big|_L = S_0(\alpha)$. For solving the eikonal equation we construct the rays, that is, the equation (18) with characteristics (rays)

$$\frac{dx}{d\sigma} = \frac{p}{K(x, y)} \qquad \frac{dp}{d\sigma} = \frac{\partial K(x, y)}{\partial x}$$

$$\qquad\qquad\qquad\qquad\qquad\qquad\qquad\qquad (18)$$

$$\frac{dy}{d\sigma} = \frac{q}{K(x, y)} \qquad \frac{dq}{d\sigma} = \frac{\partial K(x, y)}{\partial y}$$

where $p = \partial S / \partial x, \quad q = \partial S / \partial y, \quad d\sigma$ is the length element of the ray. The initial conditions $p_0$ and $q_0$ shall be defined from the system

$$p_0 \frac{\partial x_0}{\partial \alpha} + q_0 \frac{\partial y_0}{\partial \alpha} = \frac{\partial S_0}{\partial \alpha},$$

$$p_0^2 + q_0^2 = K^2(x_0(\alpha), y_0(\alpha)) \, .$$

The equations (3.1.9) and initial conditions $x_0(\alpha), y_0(\alpha), p_0(\alpha), q_0(\alpha)$ define the ray $x = x(\sigma, \alpha), y = y(\sigma, \alpha)$. After the rays are found the eikonal $S$ is defined by integration along the ray :

$$S = S_0(\alpha) + \int_0^\sigma K(x(\sigma, \alpha), \, y(\sigma, \alpha)) d\sigma$$

The function $W_0$ is defined to the accuracy of multiplication by the arbitrary function $A_0(x, y)$. We shall find $W_0$ given as: $W_0(z, x, y) = A_0(x, y) W_0^*(z, x, y)$, where $W_0^*(z, x, y)$ is the solution of the vertical spectral problem problem (17) normalized as follows

$$\int_0^H (N^2(z,x,y) - \omega^2) W_0^{*2}(z,x,y)dz = 1$$

Finally, we can obtain a following equation

$$\nabla A_0^2 \nabla S + A_0^2 \Delta S - 3 \nabla S \nabla \ln K = 0$$

This equation will be solved in characteristics of the eikonal equation (18). Using the formula for $\Delta S$ along the rays : $\Delta S = \dfrac{1}{J} \dfrac{d}{d\sigma}(JK)$, where $J(x,y)$ is the geometric ray spread, we reduce the transfer equation (19) to the following conservation law along the rays

$$\frac{d}{d\sigma}\left( \ln \frac{A_0^2(x,y) J(x,y)}{K^2(x,y)} \right) = 0$$

Note that the wave energy flash is proportional to $A_0^2 K^{-1} da$ , thus, from this equation it follows that, in this case, there survives the value equal to the wave energy flash divided by the wave vector modulus.

To proceed to studying the problem of non-harmonic wave packets evolution in a smoothly non-uniform horizontally and non-stationary stratified medium we presuppose the choice of Anzatz ("Anzatz" is the German for a solution type), which define the propagation of Airy and Fresnel internal waves with certain heuristic arguments . The Airy waves describe the features of far wave internal gravity fields in shelf zone, the Fresnel waves describe the features of far wave internal gravity fields in deep ocean.

Airy wave. Let's introduce the slow variables $x^* = \varepsilon x$, $y^* = \varepsilon y$, $t^* = \varepsilon t$ (no slowness is supposed over $z$ , the index is omitted hereafter), where $\varepsilon = \lambda / L << 1$ is the small parameter that characterizes the softness of ambient horizontal changes ( $\lambda$ is the typical iternal gravity wave length, $L$ is the scale of a horizontal non-uniformity). Next we examine the superimposition of harmonic waves (in slow variables $x, y, t$ )

$$W = \int \omega \sum_{m=0}^{\infty} (i\varepsilon)^m W_m(\omega, z, x, y) \ \exp(F) d\omega$$

$$F(x,y,t) = \frac{i}{\varepsilon}\big[\omega t - S_m(\omega, x, y)\big]$$

With respect to functions $S_m(\omega, x, y)$ it is assumed that they are odd-numbered on $\omega$ and $\min\limits_{\omega} \partial S / \partial \omega$ is reached at $\omega = 0$ (for all $x$ and $y$ ). Substituting this representation into (20) we can easily have it proved that the function $W_m(\omega, z, x, y)$ has at $\omega = 0$ a pole of the m-th order . Therefore, as the model integral $R_m(\sigma)$ for individual terms will serve the following formulas

$$R_m(\sigma) = \frac{1}{2\pi} \int\limits_{-\infty}^{\infty} \left(\frac{i}{\omega}\right)^{m-1} \exp(i\,\omega^3/3 - i\sigma\omega) d\omega$$

where the integration contour is going around the point $\omega = 0$ from overhead, which enables the functions $R_m(\sigma)$ to exponentially decay at $\sigma \gg 1$. The functions $R_m(\sigma)$ have the following feature:

$$\frac{d\,R_m(\sigma)}{d\sigma} = R_{m-1}(\sigma), \quad \text{at that} \quad R_0(\sigma) = Ai'(\sigma), \quad R_1(\sigma) = Ai(\sigma), \quad R_2(\sigma) = \int\limits_{-\infty}^{\sigma} Ai(u) du, \quad \text{etc.} \quad \text{It is evident,}$$

considering certain properties of Airy integrals , that the functions $R_m(\sigma)$ related with each other as

$$R_{-1}(\sigma) + \sigma\,R_1(\sigma) = 0$$

$$R_{-3}(\sigma) + 2\,R_0(\sigma) - \sigma^2\,R_1(\sigma) = 0 \,.$$

Fresnel wave. As the model integrals $R_m(\sigma)$ that describe the propagation of Fresnel waves taking into account the solution structure for the displacement in the horizontally uniform case we use the following formulas

$$R_0(\sigma) = \text{Re} \int\limits_{0}^{\infty} \exp\left(-it\sigma - it^2/2\right) dt \quad R_{-1}(\sigma) + i\sigma\,R_0(\sigma) = 0 \quad R_{-3}(\sigma) - 2i\,R_{-1}(\sigma) - i\sigma\,R_{-2}(\sigma) = 0 \,.$$

Based on the above, and as well on the first member structure of the Airy and Fresnel uniform wave asymptotics for a horizontally uniform medium, the solution of the system in (20) can be found, for instance, in the form (for an individually taken mode $W_n$, $\mathbf{U}_n$, further omitting the index $n$ )

$$W = \varepsilon^0 W_0(z,x,y,t) R_0(\sigma) +$$

$$+ \varepsilon^a W_1(z,x,y,t) R_1(\sigma) + \varepsilon^{2a} W_2(z,x,y,t) R_2(\sigma) + \ldots$$

$$\mathbf{U} = \varepsilon^{1-a} \mathbf{U}_0(z,x,y,t) R_1(\sigma) + + \varepsilon \mathbf{U}_1(z,x,y,t) R_2(\sigma) + \varepsilon^{1+a} \mathbf{U}_2(z,x,y,t) R_3(\sigma) + \ldots$$

where the argument $\sigma = \left(S(x,y,t)/a\right)^a \varepsilon^{-a}$ is assumed to be of the order of unity. This expansion agrees with a common approach of the geometric optics method and space-time ray-path method .

Note also that from such a solution structure it follows that the solution for a horizontally non-uniform and non-stationary medium shall depend on both the "fast" (vertical coordinate) and "slow" (time and horizontal coordinates) variables. Next we generally are going to find a solution in "slow" variables, at that the solution's structural elements which depend on the "fast" variables appear in the form of integrals of some slowly varying functions along the space-time rays.

This solution choice allows us to define the uniform asymptotics for internal gravity wave fields propagating within stratified mediums with slowly varying parameters, which holds true either near or far away from the wave fronts of a single wave mode. If we need only to define the behavior of a field near the wave front, then we can use one of the geometric optics methods – the "progressing wave" method, and a weakly dispersive approximation in the form of appropriate local asymptotics, and find the representation for the phase functions argument $\sigma$ in the form: $\sigma = \alpha(t, x, y)(S(t, x, y) - \varepsilon t) \varepsilon^{-a}$, where the function $S(t, x, y)$ defines the wave front position and is determined from the eikonal equation solution: $\nabla^2 S = c^{-2}(x, y, t)$, where $c(t, x, y)$ is the maximum group velocity of a respective wave mode, i.e., the first member of the dispersion curve expansion in zero. The function $\alpha(t, x, y)$ (the second member of the expanded dispersion curve) describes the space-time impulse width evolution of Airy or Fresnel non-harmonic internal gravity waves, and then it will be defined from some arbitrary laws of conservation along the eikonal equation characteristics with their actual form to be determined by the problem physical conditions.

## 2. NUMERICAL SIMULATION

*2.1. Wave dynamics in vertically stratified mediums.* Under consideration is the problem of mathematical modeling for the field of steady-state internal gravity waves generated by a non-local disturbing source (for example underwater sea platform) within a flow of stratified medium of the thickness $H$ with an arbitrary distribution of the Vaisala-Brunt frequency $N(z)$. The free surface at $z = 0$ is substituted with the "rigid-lid" which allows us to filter off the surface waves, and has little effect upon the internal gravity waves. It is assumed that a flow velocity $V$ exceeds the maximum group velocity of internal waves in real ocean (Arctic basin). The disturbing non-local source vertical dimension is considered small as compared to the medium layer thickness. These assumptions mean that the internal Froude number is much greater than unity, so the pictures of the trajectories near a flowing source must qualitatively appear the same as in the case of the uniform (non-stratified) medium.

Parameters of the calculations are typical for Arctic basin and underwater sea construction: $N(z) \approx 0.01\,s^{-1}$, sea depth $H \approx 100\,m$, stratified flow velocity $V \approx 2m/s$, $\xi = x + Vt$, horizontal scale of underwater streamlined obstacle is about 50 m. The numerical simulation of the problem stated requires quite a number of integrations from the fast oscillating functions, thus; first, we have to use methods which allow us to effectively realize the integrations of this type. Second, the complete wave field near a non-local flowing source of perturbations represents a poorly convergent series, and to obtain an adequate accuracy we have to integrate a large number of modes, however, the use of the static feature discrimination method enables the calculation of the field near a flowing source while avoiding such integrations. Finally, third, at long distances from the source when the complete field falls into singular modes, the asymptotic representations for a single mode of the Green's function, make it possible to calculate the far fields of internal gravity waves without performing exact numerical calculations.

  The numerical calculation show, for example, that vertical velocity W is quickly decreasing with decreasing depth and at $z = Z$ ($z = Z$ - depth of the thermocline maximum) it takes about 15% of the velocity value at the bottom. Fig.1 demonstrate the calculation results in the thermocline maximum (integrated were 25 wave modes, the higher modes had not contributed much to the complete field), the maximum value of the displacement at this horizon reached 1.3 meters. The presented results show that there are at least three different regions of the generated field of internal waves. First, it is the region immediately under the non-local source, which has a width of about the medium thickness, it's the near-field region. The numerical calculations have proven that the wave field of internal gravity waves within the near-field region is little dependent on a specific stratification and the velocity amplitude and displacement within this region are maximal. Secondly, at long distances from the non-local source ($y, x > 10H$, the far-field region) the field of internal gravity waves falls apart into singular wave modes, at that each of the modes is contained inside its Mach cone, and outside the cline the amplitude is low. In addition to that there is a transition region in which the structure of the wave fields is rather complex.

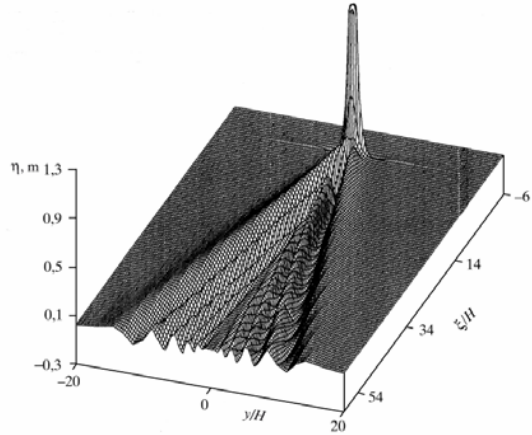

**Fig. 1:** *Internal gravity waves (vertical displacement) from a underwater nol-local source in stratified medium of uniform depth* ($\xi = (x + Vt) / H, y = y / H$ *- non-dimensional horizontal coordinates*)

*2.2 Wave dynamics in horizontally inhomogeneous mediums.* In Fig. 2 we represent vertical component of internal gravity wave field velocity $W$ generated by a non-local source (underwater obstacle – sea platform) in arbitrary stratified ocean of non-uniform depth. Parameters of the calculations are typical for Arctic basin: $N(z) \approx 0.001 \, s^{-1}$, the slope of the bottom no more than $10^0$. Numerical calculations show a significant deformation of the wave field structure, taking into account the horizontal inhomogeneities stratified mediums. For example, it follows from the numerical results thus presented that, outside the caustic, the wave field is sufficiently small indeed and is not subjected to great many oscillations, whereas the wave picture inside the zone of caustic is a rather complicated system of incident and reflected harmonics. It is well known, that caustic is an envelope of a family of rays, and asymptotic solution is obtained along these rays. Asymptotic representation of the field describe qualitative change of the wave field, and that is description of the field, when we cross the area of "light", where wave field exists, and come in the area of "shadow", where we consider wave field to be rather small. Each point of the caustic corresponds to a specified ray, and that ray is tangent at this point. In this paper the most difficult question is considered that can appear when we investigate the problems of wave theory with the help of geometrical optics methods and its modifications. And the main question consists in finding of asymptotic solution near special curve (or surface), which is called caustic.

It is a general rule that caustic of a family of rays single out an area in space, so that rays of that family cannot appear in the marked area. There is also another area, and each point of that area has two rays that pass through this point. One of those rays has already passed this point, and another is going to pass the point. Formal approximation of geometrical optics or WKBJ approximation cannot be applied near the caustic, that is because rays merge together in that area, after they were reflected by caustic. If we want to find wave field near the caustic, then it is necessary to use special approximation of the solution, and in the paper a modified ray method is proposed in order to build uniform asymptotic expansion of integral forms of the internal gravity wave field. After the rays are reflected by the caustic, there appears a phase shift. It is clear that the phase shift can only happen in the area where methods of geometrical optics, which were used in previous sections, can't be applied. If the rays touch the caustic several times, then additional phase shifts will be added. Phase shift, which was created by the caustic, is rather small in comparison with the change in phase along the ray, but this shift can considerably affect interference pattern of the wave field.

The asymptotic representations constructed in this paper allow one to describe the far field of the internal gravity waves generated by a non-local sources in stratified flow. The obtained asymptotic expressions for the solution are uniform and reproduce fairly well the essential features of wave fields near caustic surfaces and wave fronts. In this paper the problem of reconstructing non-harmonic wave packets of internal gravity waves generated by a source moving in a horizontally stratified medium is considered. The solution is proposed in terms of modes, propagating independently in the adiabatic approximation, and described as a non-integer power series of a small parameter characterizing the stratified medium. In this study we analyze the evolution of non-harmonic wave packets of internal gravity waves generated by a moving source under the assumption that the parameters of a vertically stratified medium (e.g. an ocean) vary slowly in the horizontal direction, as compared to the characteristic length of the density. A specific form of the wave packets, which can be parameterized in terms of model functions, e.g. Airy functions, depends on local behavior of the dispersion curves of individual modes in the vicinity of the corresponding critical points.

In this paper a modified space-time ray method is proposed, which belongs to the class of geometrical optics methods. The key point of the proposed technique is the possibility to derive the asymptotic representation of the solution in terms of a non-integer power series of the small parameter $\varepsilon = \lambda / L$, where $\lambda$ is the characteristic wave length, and L is the characteristic scale of the horizontal heterogeneity. The explicit form of the asymptotic solution was determined based on the principles of locality and asymptotic behavior of the solution in the case of a stationary and horizontally homogeneous medium. The wave packet amplitudes are determined from the energy conservation laws along the characteristic curves. A typical assumption made in studies on the internal wave evolution in stratified media is that the wave packets are locally harmonic. A modification of the geometrical optics method, based on an expansion of the solution in model functions, allows one to describe the wave field structure both far from and at the vicinity of the wave front.

Using the asymptotic representation of the wave field at a large distance from a non-local source in a layer of constant depth, we solve the problem of constructing the uniform asymptotics of the internal waves in a medium of varying depth. The solution is obtained by modifying the previously proposed "vertical modes-horizontal rays" method, which avoids the assumption that the medium parameters vary slowly in the vertical direction. The solution is parameterized, for example, through the Airy waves. This allows one to describe not only the evolution of the non-harmonic wave packets propagating over a slow-varying fluid bottom, but also specify the wave field structure associated with an individual mode both far from and close to the wave front of the mode. The Airy function argument is determined by solving the corresponding eikonal equations and finding vertical spectra of the internal gravity waves. The wave field amplitude is determined using the energy conservation law, or another adiabatic invariant, characterizing wave propagation along the characteristic curves.

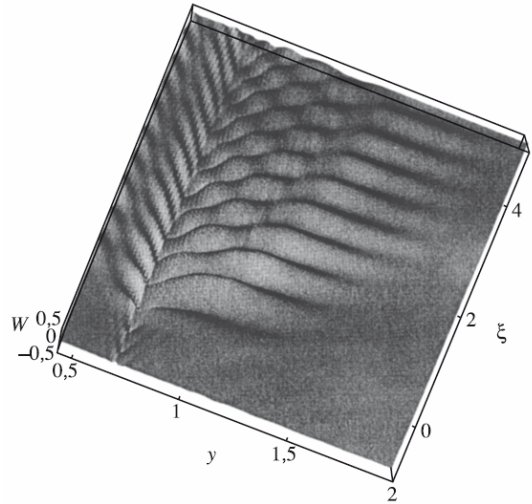

**Fig. 2: *Internal gravity waves (vertical velocity component) from a underwater nol-local source in stratified medium of variable depth ( $\xi = (x+Vt)/H$, $y = y/H$ - non-dimensional horizontal coordinates)***

Modeling typical shapes and stratification of the ocean shelf in Arctic basin we obtain analytic expressions describing the characteristic curves and examine characteristic properties of the wave field phase structure. As a result it is possible to observe some peculiarities in the wave field structure, depending on the shape of ocean bottom, water stratification and the trajectory of a moving source. In particular, we analyze a spatial blocking effect of the low-frequency components of the wave field, generated by a source moving alongshore with a supercritical velocity. Numerical analyses that are performed using typical ocean parameters reveal that actual dynamics of the internal gravity waves are strongly influenced by horizontal non-homogeneity of the ocean bottom. In this paper we use an analytical approach, which avoids the numerical calculation widely used in analysis of internal gravity wave dynamics in stratified ocean (Arctic basin).

**CONCLUSIONS**

The main fundamental problems of wave dynamics considered in the present paper were the following:

  - construction of the exact and asymptotic solutions of the problem concerning the  internal gravity waves excited by the non-local disturbing sources in the non-uniform stratified mediums, as well as development of the numerical algorithms for analysis of the corresponding spectral problems and for calculation of the wave disturbances for the real parameters of the vertically stratified mediums;

  -  research by means of the modified version of the space-time ray-tracing method (WKBJ method), evolution of the non-harmonic wave-trains of the internal gravity waves in the supposition of the slowness of variation of the parameters of the vertically stratified medium in the horizontal direction and in a time;

  - the asymptotic analysis of the critical modes of generation and propagation of the internal gravity waves in the stratified mediums, including the study of the effects of the space-frequency screening;

  - development of non-spectral methods of analysis of the in-situ measurements of the internal gravity waves for the purpose of the possible distant definition of the characteristics of the broad-band wave-trains, composing the measured hydrophysical fields, as well as the parameters of the ocean along a line of propagation of these wave-trains.

        The paper presented  methods and approaches of research of the internal gravity waves dynamics combine the comparative simplicity and computational capability to gain the analytical results, the possibility of their qualitative analysis and the accuracy of the numerical results. Besides that there is a possibility of inspection of the trustworthiness of the used hypotheses and approximations on the basis of analysis of the real oceanological data, while the exact analytical solutions for the model problems do not allow to apply the gained outcomes, for example, for analysis of the problem with  the real parameters of the medium, and the exact numeric calculation for one particular real medium does not give the possibility of the qualitative analysis of the medium with other real parameters.

      The results presented by the paper on the research of the dynamics of the non-harmonic wave-trains of the internal waves in the stratified mediums with the varying parameters enable analytically and numerically to examine effects of the special blocking, and also the excitation and failure of the separate frequency components of the propagating wave-trains.

        It is necessary to mark once again, that in comparison with the majority of the researches devoted to study of the dynamics of the internal gravity waves, the methods of decomposing of the fields of the internal gravity waves into the certain benchmark functions enable to describe the main peculiarities of formation of the critical modes of generation and propagation of the non-harmonic wave-trains. It is expedient also to emphasize, that the built asymptotic representations in the form of the applicable model functions can be used also for study of any other wave processes (acoustical and seismic waves, SHF-irradiation, the tsunami waves, etc.) in  the real mediums with a complex structure. All fundamental results of the paper are gained for the arbitrary distributions

of the density and other parameters of the non-uniform media, and besides the main physical mechanisms of formation of the studied phenomena of the dynamics of the internal gravity waves in the non-uniform stratified mediums were considered in the context of the available data of the in-situ measurements.

The next step in the asymptotic study of the internal gravity waves should be study of the linear interaction of the wave-trains at their propagation as we used approximation of adiabatic, that is the independence of wave modes from each other. However, generally, the linear interaction ( the linear conversion) of the waver modes is present. The phenomenon of the linear conversion of the internal gravity waves consists, that at the wave-trains passing through the non-uniform sections of the medium the amplitudes of the waves can vary non-adiabatically, that is the real amplitude-phase characteristics of the fields are varying differently, than it follows from the fundamental approximations of the geometrical optics used in this paper. The detailed study of these problems will be the subject of further researches.

The universal nature of the he asymptotic methods of research of the internal gravity waves offered in this paper is added with the universal heuristic requirements of the applicability of these methods. These criteria ensure the internal control of applicability of the used methods, and in some cases on the basis of the formulated criteria it is possible to evaluate the wave fields in the place, where the given methods are inapplicable. Thus there are the wide opportunities of analysis of the wave patterns as a whole, that is relevant both for the correct formulation of the analytical investigations, and for realization of estimate calculations at the in-situ measurements of the wave fields.

The special role of the given methods is caused by that condition, that the parameters of the natural stratified mediums, as a rule, are known approximately, and efforts of the exact numerical solution of initial equations with usage of such parameters can lead to the overstatement of accuracy.

Also popularity of the used approaches of analysis of dynamics of the internal gravity waves can be promoted just by the existence of the lot of the interesting physical problems quite adequately described by these approaches and can promote the interest to the multiplicity of problems bound to a diversification of the non-uniform stratified mediums. The value of such methods of analysis of the wave fields is determined not only by their obviousness, scalability and effectiveness at the solution of the different problems, but also that they can be some semi-empirical basis for other approximate methods in theory of propagation of the internal gravity waves.

The results of this paper represent significant interest for physics and mathematics. Besides, asymptotic solutions, which are obtained in this paper, can be of significant importance for engineering applications, since the method of geometrical optics, which we modified in order to calculate the wave field near caustic, makes it possible to describe different wave fields in a rather wide class of other problems

**APPLICATIONS**

Industrial activities on the continental and Arctic shelf connected with oil, gas, and other minerals extraction became one of the important reasons to begin researches of dynamic internal gravity waves. Ships and platforms busy with drilling and construction at the depth use long tubes joining them with the sea bottom. Builders of underwater constructions in equatorial districts experienced the influence of huge underwater internal waves and strong surface flows which can have the form of steep waterfalls. Some time ago when the phenomenon of internal waves and their strength were not known it happened that the builders lost their equipment. Such expensive losses made them think that security of underwater equipment and the influence of internal gravity waves should be controlled.

Construction of sea platforms, such as Ormen Lange (Norway), and others on the Arctic shelf stimulated numerous scientific researches, including fundamental. Thermocline at the Ormen Lange deposit lies at the depth of about 500 m. It separates warm Atlantic water (about 7 Centigrade) from cold polar water (about 1 Centigrade). Additional accumulation of warm Atlantic stream can put the thermocline lower. Measurements near Ormen Lange (Norway) fixed once that the stream moved the thermocline from the usual depth to the depth of 550 m, and it stayed there for 3 days. The thermocline reached the platform at the depth of 850 m. Later water moved backward and upward along the slope. First its velocity was half a meter per second which is too small for the bottom stream. Step by step it slowered but oscillations went on for a long time – for the whole day.

The amplitude of internal gravity waves is usually comparable with the depth of the ocean surface layer, but it was fixed when the wave was 5 times higher than the thermocline. As the sea water consists of layers, one above another, and having different temperature and salinity, internal gravity waves exist at all depths in the ocean and reach their maximal amplitudes as a rule near the thermocline. In equatorial districts thermocline is situated at the depth of 200-300 m, in the districts of oil, gas Arctic deposits – about 500 m, and in Norwegian fjords with coming fresh water – at the depth only 4-10 m.

The internal waves characteristics are used for appreciation of their influence on the environment and underwater platforms of oil and gas deposits at the Arctic shelf. Stationary tubes for oil and gas transportation stretch along the ocean shelf slope. These tubs are about 500 m long and they can suffer from internal gravity waves. That is why calculations of wave dynamics are used for appreciation of the sea platform equipment wear.

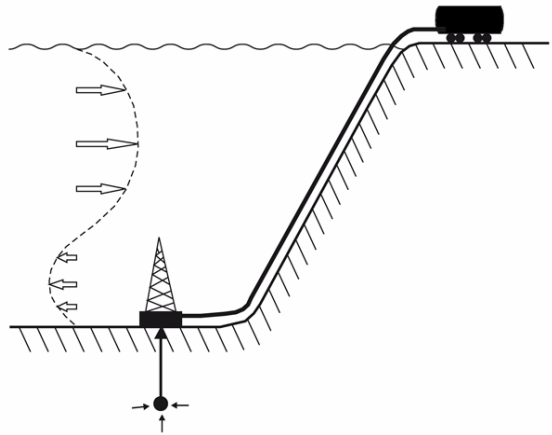

**Fig. 3:** *Gas (oil) production in the Arctic Basin. Gas (oil) deposits are located at a depth of several hundred meters and distances of ten or more kilometers from the coast. Internal gravity waves affect the underwater technical objects.*

Internal waves play the role of transport moving biomass and nutrient matters from place to place. Gliding upwards along the shelf they bring nutrient matters from the depth to more salted shoal, where conditions for life of fries and larvae are ideal. The internal waves movement in this case can be compared with the work of a pump. There is an interesting connection between internal waves and the sea life. In slow and long vertical stream formed by these waves plankton and small sea organisms can live. Experiments show that sea organisms use such vertical streams. They can swim vertically against the current and grow and propagate at the same time. Such processes take place just along the vertical stream while moving of the wave. They are observed with the help of sputnics in the Arctic districts rich with fish resources, for example, in straits between the Kara and Barents seas.

Special interest to research of internal gravity waves is connected with also intensive exploitation of Arctic and its natural wealth. Internal waves in Arctic are poorly studied as they move under ice and practically invisible from above, but accessible information about underwater objects movement show their existence. Sometimes there are exclusions when internal gravity waves reach ice and uplift and lower it with definite periodicity which can be fixed with the help of radiolocation sounding. Influence of all kinds of waves can be the reason of the ice cover split in the Arctic. Internal waves make for the movement of icebergs and different kinds of pollution. So, the research of wave dynamics in the region of the Arctic shelf is an important fundamental scientific and practical problem aimed at ensuring security while.

**ACKNOWLEDGMENTS**

The results presented in the paper have been obtained by research performed under projects supported by the Russian Foundation for Basic Research (No.11-01-00335a "Stability and Instability of Flows with Interfaces" , No. 12-05-00227a "Dynamics of Wave Processes in Conditions of the Arctic Basin"), Program of the Russian Academy of Sciences "Fundamental Problems of Oceanology: Physics, Geology, Biology, Ecology" .